\newcommand{\gcm}{{\rm g\,cm^{-2}}}
\newcommand{\Xmax}{X_{\rm max}}
\newcommand{\Xmumax}{X^{\mu}_{\rm max}}
\newcommand{\thmax}{\Theta_{\rm max}}
\newcommand{\lesim}{^{_<}_{^\sim}}

\documentclass[a4paper]{jpconf}

\usepackage{graphicx}
\usepackage{subfigure}          
\begin{document}
\title{Studying the nuclear mass composition of Ultra-High Energy Cosmic Rays with the Pierre Auger Observatory}
\author{L.~Cazon$^1$ for The Pierre Auger Collaboration$^{2,3}$}
\address{$^1$ LIP, Av Elias Garcia 14-1,1000-149 Lisboa, Portugal}
\address{$^2$ Av. San Mart\' \i n Norte 304 (5613) Malarg\"ue, Argentina}
\address{$^3$ A full author list and affiliations can be found at http://www.auger.org/archive/authors\_2011\_09.html}

\ead{cazon@lip.pt}
 
\begin{abstract}

The Fluorescence Detector of the Pierre Auger Observatory measures the atmospheric depth, $\Xmax$, where the longitudinal profile of the  high energy air showers reaches its maximum. This is sensitive to the nuclear mass composition of the cosmic rays. Due to its hybrid design, the Pierre Auger Observatory also provides independent experimental observables obtained from the Surface Detector for the study of the nuclear mass composition. We present $\Xmax$-distributions and an update of the average and RMS values in different energy bins and compare them to the predictions for different nuclear masses of the primary particles and hadronic interaction models. We also present the results of the composition-sensitive parameters derived from the ground level component.

\end{abstract}

\section{Introduction}
The determination of the nuclear mass composition of UHECR (Ultra High Energy Cosmic Rays) is fundamental to unveil the origin of the most energetic particles known in nature. UHECR are detected by means of the extensive air showers created in the Earth's atmosphere,  which are composed  by a cascade of hadrons  and electromagnetic (EM) particles. The depth at which the EM shower reaches its maximum, $\Xmax$, strongly correlates with the depth where the primary firstly interacted. The $\Xmax$-distribution  carries information about the primary particle and the physical processes in the cascade. The hadronic cascade is composed mostly by pions, of which the charged pions might decay in flight into a muon and a neutrino. Due to their long lifetime and low cross section, muons can leave the core of the hadronic cascade traveling kilometers away and be detected.  Electrons spread out in time and space because of Coulomb scattering, whereas muons practically travel following straight lines.  The time structure of the shower disc encodes the history of the shower, allowing us to recover information about the longitudinal evolution by analyzing the time distribution of the particles arriving at ground level.

The Pierre Auger Observatory, located on the high plateau of the Pampa Amarilla, is the largest cosmic ray observatory ever built. Its hybrid design allows to collect the shower particles by a surface detector (SD) and to observe the longitudinal development of the EM profile by collecting the UV light with a fluorescence detector (FD). The SD  spans 1600 Cherenkov detectors in a 1.5 km triangular grid
 over 3000 km$^2$, whereas the FD is composed by 24 telescopes distributed over 4 sites overlooking the array. The baseline design is being complemented with enhancements both on the SD and FD, and new detection techniques \cite{enhancements}. 

In this paper we present the latest $\langle\Xmax \rangle$ and RMS-$\Xmax$ results as a function of energy, and the $\Xmax$-distributions for the highest and lowest energy bins. We also present the evolution with energy of SD observables extracted from the time distributions of the signal which are sensitive to the longitudinal evolution of the shower. We compare our results with the predictions of hadronic interaction models \cite{Models}.

\section{Electromagnetic Shower Maximum}

We have used data taken between December 2004 and September 2010,  following the analysis reported in \cite{Abraham:2010yv} and updated in \cite{Facal:2011}. We have considered those showers reconstructed by the FD that have at least a signal in one of the SD stations. The longitudinal profile was fitted with a Gaisser-Hillas function, and a series of cuts depending on the atmospheric optical conditions and on the quality of the reconstructed profile  were applied \cite{Abraham:2010yv,Facal:2011}. In order to avoid that the data sample is biased regarding to the nuclear mass, a number of cuts in the geometry were imposed.

\begin{figure}[h] \begin{center}
\begin{minipage}{7cm}
\includegraphics[width=7cm]{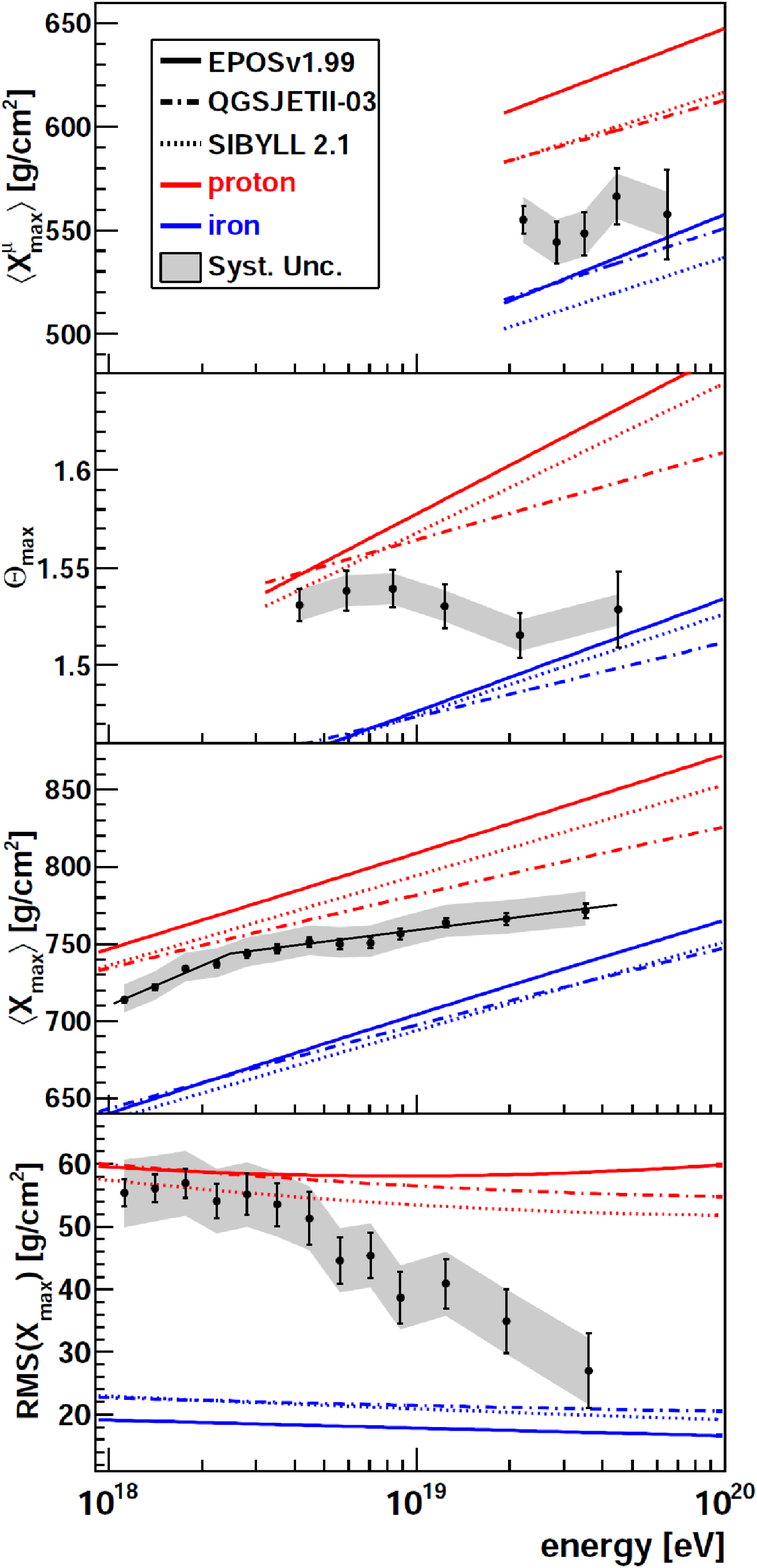}
\end{minipage}\hspace{2pc}%
\begin{minipage}{4.7cm} 
\includegraphics[width=4.7cm]{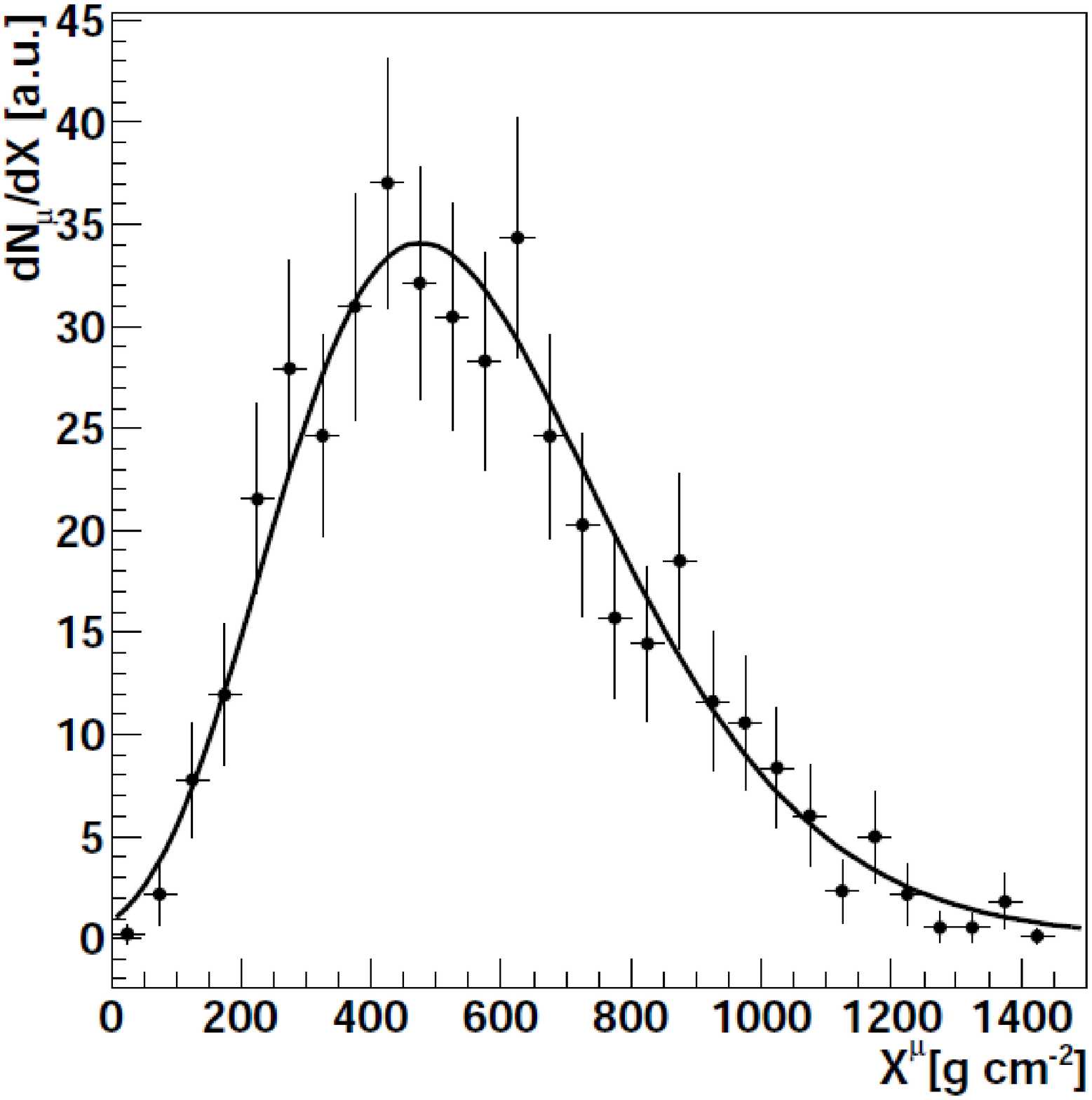}\\
    \includegraphics[width=4.7cm]{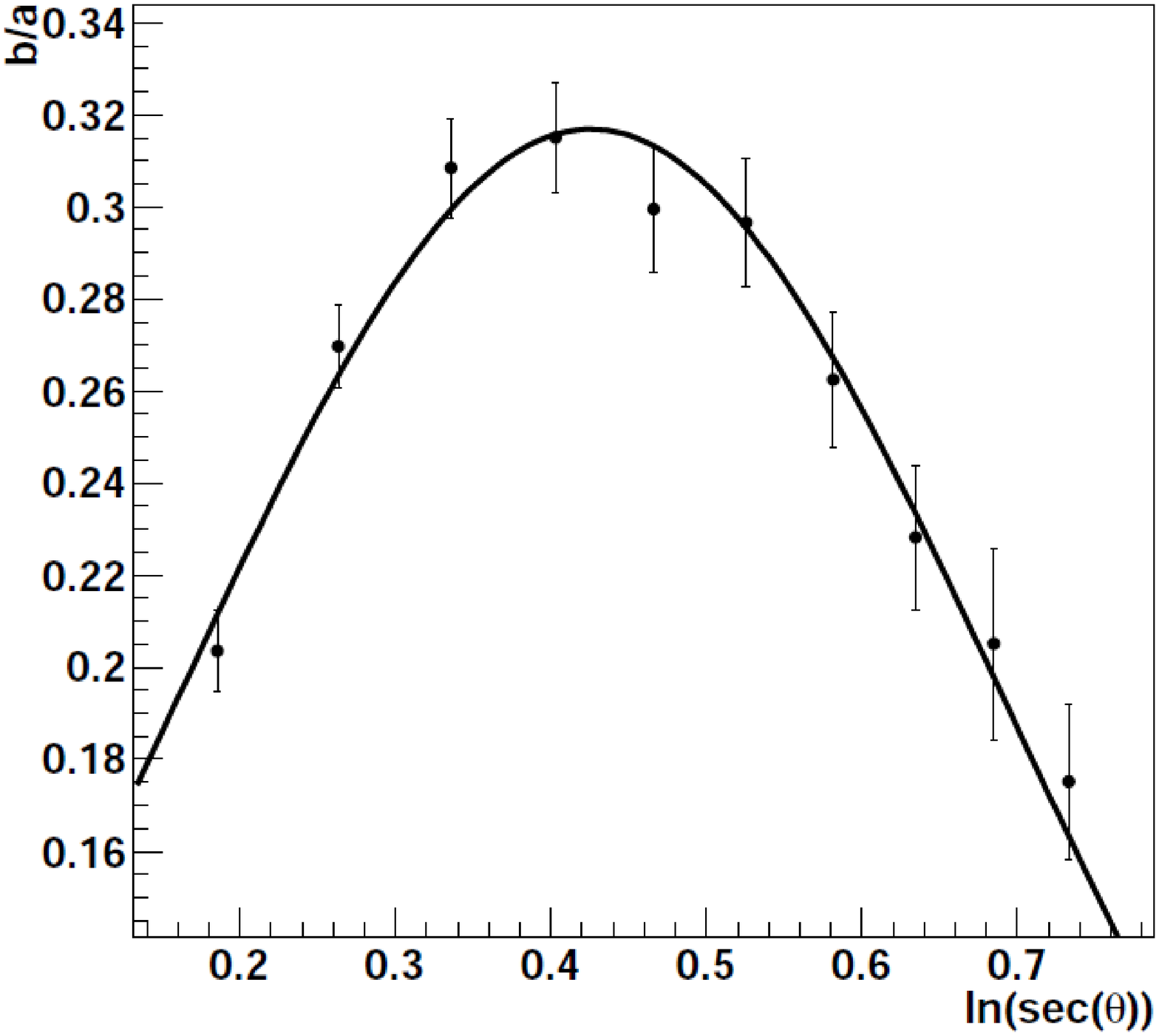}  \\
 \includegraphics[width=4.7cm]{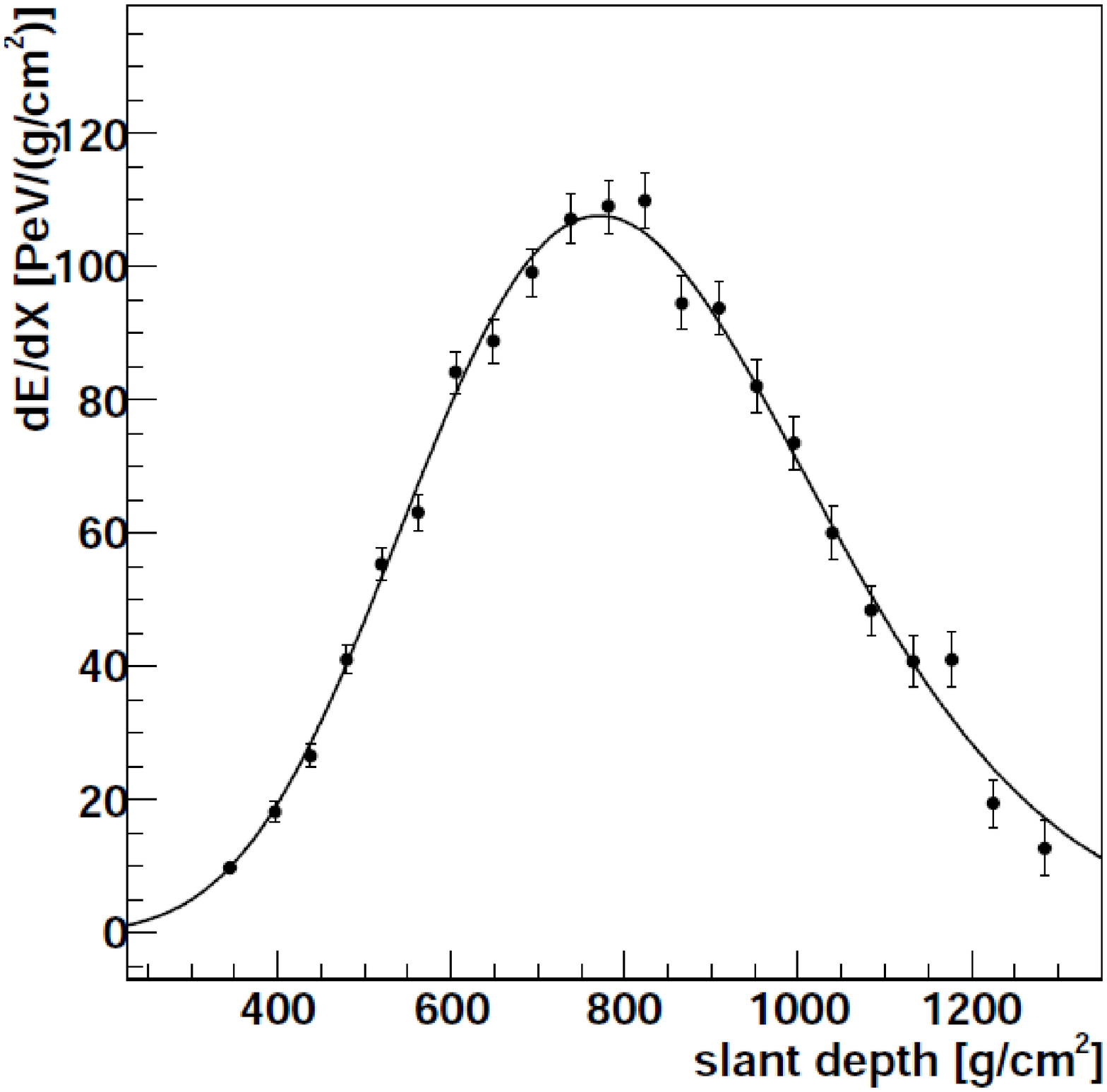} 
\end{minipage} 
  \caption[]{Left Panel: Results on shower evolution sensitive observables compared with models predictions. The error bars correspond to the statistical uncertainty, whereas the systematic uncertainty is represented by the shaded bands. Right Panel, top: Typical longitudinal development of the muon production. Middle: Average asymmetry in the risetime. Bottom: Typical longitudinal development on the energy deposit.}
    \label{EM_Assym}
  \end{center}
\end{figure}

The resolution in $\Xmax$ is around 20 $\gcm$. Uncertainties in the atmospheric conditions, calibration, event selection and reconstruction result in a systematic uncertainty of $\leq  13$ $\gcm$. The average values as a function of the energy are displayed in the left panel of Fig.~\ref{EM_Assym} (third plot from top), alongside with the predictions for different hadronic interaction models.

\begin{figure}[h]
\begin{center}
\begin{minipage}{7cm}
\includegraphics[width=7cm]{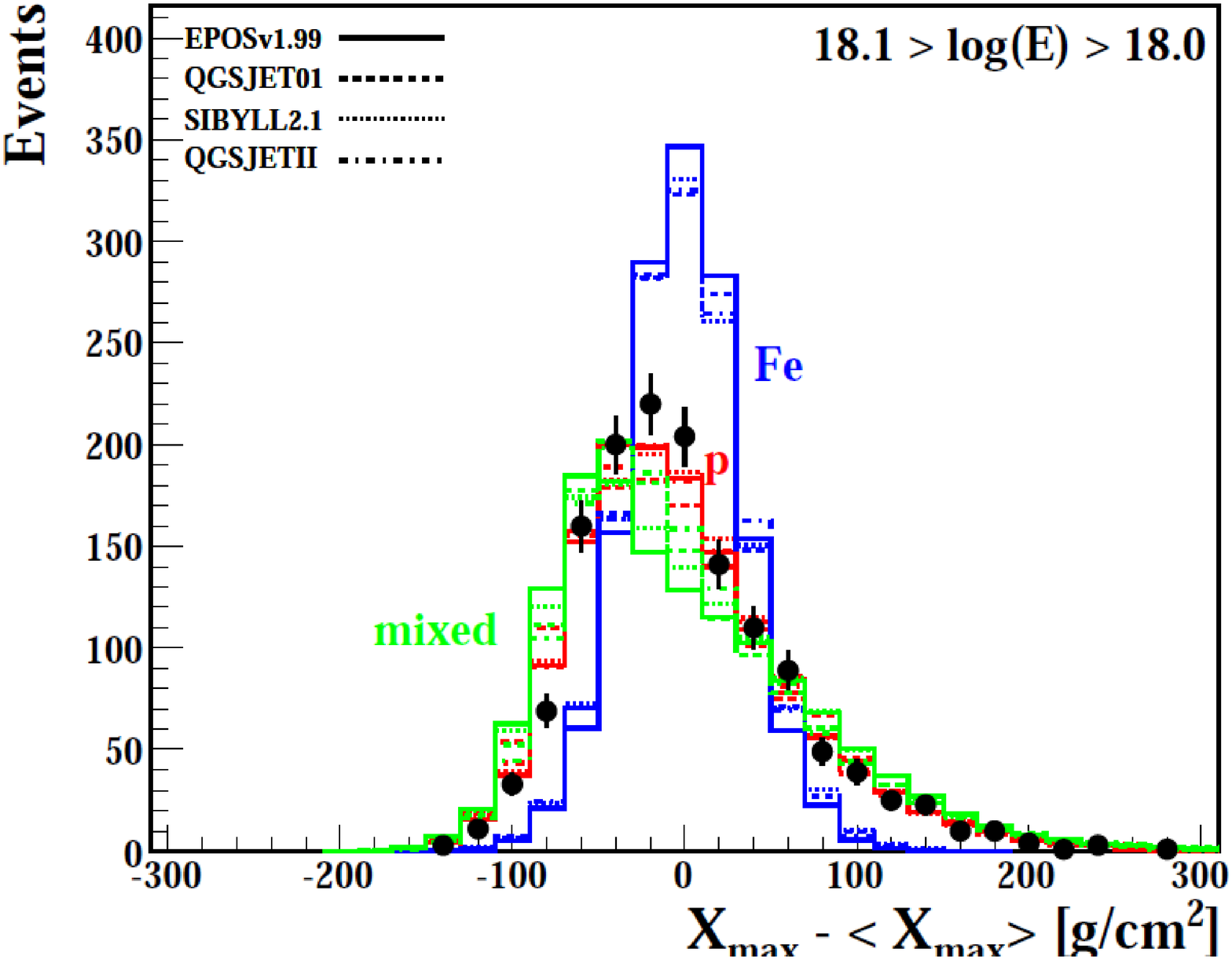}
\end{minipage}\hspace{2pc}%
\begin{minipage}{7cm}
\includegraphics[width=7cm]{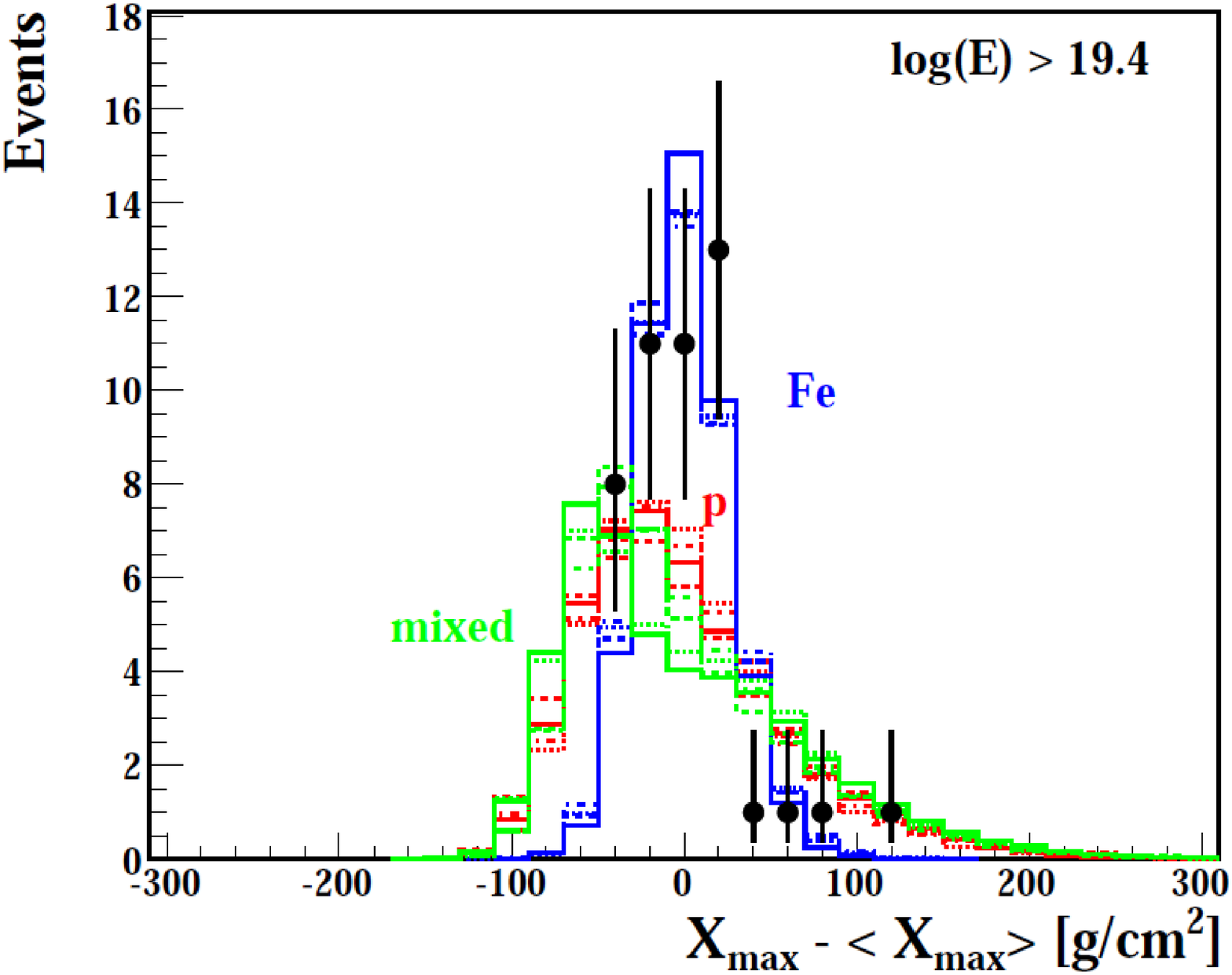}
\end{minipage} 
 \caption[]{Centered distributions $\Xmax-\langle \Xmax \rangle$, for the lower and highest energy bins. The mixed composition corresponds to 50\% p and Fe.}
 \label{Xmaxmodels}
\end{center} 
\end{figure}

The elongation rate $D_{10}=d \langle \Xmax \rangle / d \log_{10}E$ is better described using two slopes, with $\chi^2/ndf$=7.4/9, compared to 54/11 if we used a single power law. The change of the elongation rate could be interpreted as a change on the nuclear mass composition, provided that there is no change on fundamental properties of the hadronic physics.  The comparison of $\langle \Xmax \rangle$ with the different predictions of the hadronic interaction models could be interpreted as a change of composition towards heavier elements at the highest energies. The left panel of Fig.~\ref{EM_Assym} also displays the RMS-$\Xmax$, which has been corrected by the detector resolution. It rapidly decreases with energy around  the same energy of the change on the elongation rate.

The distributions of $\Xmax$ for all energy bins have been published in \cite{Facal:2011}. Fig.~\ref{Xmaxmodels} plots the $(\Xmax-\langle \Xmax \rangle)$-distributions for the highest and lowest energy bins. The predictions of the different hadronic models show a nearly universal shape. At low energies, the shape of the data distribution is compatible with a very light or mixed composition, whereas at high energies a heavier composition is favored. The high $\Xmax$ tail, characteristic of very light components, is suppressed with respect to the lowest energies.

\section{Asymmetry of the signal risetime}
In each SD event, the Cherenkov detectors record the signal as a function of time. The risetime, $t_{1/2}$, defined as the time elapsed between the 10\% and 50\% of the integrated signal, depends on the distance to the shower maximum \cite{Wahlberg:2009}, the zenith angle $\theta$ and the distance to the core $r$.  In inclined showers, the average risetime  depends on an angle defined on the perpendicular plane $\zeta$ as $\langle t_{1/2}/r \rangle=a+b\cos \zeta$, due to the different effective distances of the tanks to the shower maximum.  The evolution of $b/a$ with zenith angle reaches a maximum at $\thmax$ which is different for different primary masses \cite{Dova:2009az}. In the right central panel of Fig.~\ref{EM_Assym} an example of $b/a$ as a function of $\ln(\sec\theta)$ is shown for the energy bin $\log_{10} (E/eV)=18.85- 19.00$. 

Events with energy above $3.16 \times 10^{18}$ eV and $\theta \leq 60^{\circ}$ were selected \cite{GarciaPinto:2011}.  Results of $\thmax$ as a function of energy are displayed in the left panel of Fig.~\ref{EM_Assym}. The systematic uncertainty amounts to $\lesim$ 10\% of the proton-iron separation predicted by models. The number of muons predicted by the hadronic interaction models differs from data \cite{Allen:2011}.  A preliminary study including this effect indicates a possible change of about $\leq 5$\% of the proton-iron difference.

\section{The muon production depth}

An approximated relation between the arrival time delay of a muon with respect to the shower front plane and the distance of the muon production point to ground can be used to transform the arrival time distributions of muons into production depth distributions \cite{Cazon:2004zx}. In \cite{GarciaGamez:2011}, this technique was applied to Auger data in an angular window between $55^{\circ}$ and $65^{\circ}$ for stations with $r\geq 1800$ m, as a result of a trade off between EM contamination rejection and the intrinsic resolution of the method. In Fig.~\ref{EM_Assym}, right top panel, an example for a real event at $E=94\pm3$ EeV is shown.  Similarly to the EM component, the depth at which the muon production rate reaches a maximum, $\Xmumax$ is a good indicator of the primary mass composition. 
After a series of cuts \cite{GarciaGamez:2011}, the measured values of $\langle \Xmumax \rangle$ are presented in the upper panel of Fig.~\ref{EM_Assym}. The systematic uncertainty due to reconstruction bias, core position, rejection of the EM component and quality cuts amounts to 11 $\gcm$, corresponding to 14\% of the proton-iron separation. Note that the discrepancy on the total number of muons observed between models and data does not affect this result, but only the normalization of the muon production depth distribution. 

\section{Conclusions}
Provided that the present models give a fair description of the physical processes and their systematics, the average primary composition of UHECR could be inferred from the left panel of Fig.~\ref{EM_Assym}, showing a change towards heavier mass composition at high energies. The evolution of $\langle \Xmax \rangle$,  $\thmax$ , and  $\langle \Xmumax \rangle$ is similar in the overlapping energy regions, despite the fact of having completely independent systematics. RMS-$\Xmax$ evolution can be accommodated with a variety of compositions since it is influenced by the shower-to-shower fluctuations and the relative differences on $\langle \Xmax \rangle$. The analysis of the shape of the $\Xmax$-distributions also points to the direction of heavier composition as we increase the energy. On the other hand, it is not guaranteed that current hadronic interaction models describe well particle physics at these energies \cite{Allen:2011}, and a different interpretation would therefore apply.

\end{document}